\documentstyle[11pt,aaspp4]{article}

\received{}
\accepted{}

\slugcomment{Accepted for publication in the Astrophysical Journal Letters}

\lefthead{Lutz, Veilleux, and Genzel}
\righthead{Mid-infrared and optical spectroscopy of ULIRGs}

\begin{document}

\title{Mid-infrared and optical spectroscopy of ultraluminous
  infrared galaxies: A comparison
   \footnote{Based on observations with ISO, an
   ESA project with instruments funded by ESA member states,
   with the participation of ISAS and NASA}
 }
\author{D. Lutz\footnote{MPE, Postfach 1603, 85740 Garching, Germany; 
   lutz@mpe.mpg.de},
  S. Veilleux\footnote{University of Maryland,
   College Park, MD 20742, USA},
  R. Genzel$^2$
  }

\begin{abstract}

New tools from Infrared Space Observatory (ISO) mid-infrared spectroscopy 
have recently become available to determine the 
power sources of dust-obscured ultraluminous infrared galaxies (ULIRGs). 
We compare ISO classifications -- starburst or active galactic 
nucleus (AGN) -- with classifications from optical spectroscopy, and 
with optical/near-infrared searches for hidden broad-line regions. 
The agreement between
mid-infrared and optical classification is excellent if optical
LINER spectra are assigned to the starburst group. The starburst nature of 
ULIRG LINERs strongly supports the 
suggestion that LINER spectra in infrared-selected
galaxies, rather than being an expression of the AGN phenomenon,
are due to shocks that are probably related to galactic superwinds.
Differences between ISO and optical classification provide clues on the 
evolution
of ULIRGs and on the configuration of obscuring dust. 
We find few ISO AGN with optical \ion{H}{2} or LINER identification,
suggesting that highly obscured AGN exist
but are not typical for the ULIRG phenomenon in general. Rather, our results
indicate that strong AGN activity, once triggered, quickly breaks
the obscuring screen at least in certain directions, thus becoming detectable
over a wide wavelength range.

\end{abstract}


\section{Introduction}

Dust obscuration is a significant concern in studies of ultraluminous 
infrared galaxies (ULIRGs, L$_{IR}>10^{12}L_{\sun}$, see \cite{sanders96} 
for a review). Evidence for both starburst and AGN activity has been
found, but the extent to which optical and
near-infrared spectroscopy probe the true power sources has remained
questionable due to the concentration of dust and gas in the
nuclear region of ULIRGs. The underlying problem is highlighted in
near-infrared spectra of ULIRGs (e.g. \cite{goldader95}). Qualitatively, their 
majority is starburst like, but quantitatively the starburst
activity traced tends to be insufficient to power the 
bolometric luminosity of these ULIRGs. This leaves a large unaccounted power
source which might be AGN or starburst.

Mid-infrared spectroscopy with ISO is able of penetrating significantly
higher dust columns of the equivalent of a visual screen extinction of
A$_V\sim50$. Using as diagnostics both mid-infrared fine structure lines and 
the mid-infrared emission features usually ascribed to polycyclic aromatic 
hydrocarbons (PAHs), Genzel et al. (1998) found that the majority of a sample
of 15 ULIRGs is predominantly starburst powered. From these data and after
correction of the mid-infrared line fluxes for extinction, the highly
obscured star forming regions are found able of powering the bulk of the
luminosity. This leaves little room for additional hidden power sources.
The recent application of the mid-infrared PAH method to larger ULIRG samples
allows a direct comparison of optical and mid-infrared diagnostics which
is the aim of this letter. 

\section{Mid-infrared and optical spectroscopy}

The ISO mid-infrared spectroscopy sample of ULIRGs consists mainly of the 
SWS and ISOPHOT-S 
spectroscopy of Genzel et al. (1998) and Lutz et al. (1998) and a recent 
addition of higher luminosity sources observed with ISOCAM-CVF for
the ZZULIRG consortium (Tran et al., in prep.). We have 
added observations of two hyperluminous infrared galaxies
(\cite{taniguchi97}; \cite{aussel98}). 
The entire mid-infrared emission from an ULIRG is included in the
24\arcsec$\times$24\arcsec\ aperture of ISOPHOT-S or the ISOCAM-CVF 
datacubes of 3\arcmin$\times$3\arcmin\ extent.
Our spectral classification relies mainly on the 
feature-to-continuum ratio of the 7.7$\mu$m PAH feature. Groundbased
observations first demonstrated
that PAH features are strong in starburst galaxies but weak or
absent in classical AGNs.
ISO spectroscopy further demonstrated the
anti-correlation between feature strength relative to the continuum
and the ionization state of the gas (\cite{genzel98}). 
Following Lutz et al. (1998), we classify mid-infrared spectra as AGN-like
if the 7.7$\mu$m PAH feature-to-continuum ratio (L/C) is less than 1, and 
starburst-like (`SB') if it is greater than 1. This is motivated by the fact 
that, with the exception of the dwarfs NGC 5253 (0.5) and NGC 6764 (1.3) all 11
template starbursts had a L/C of more than 1.9, with a median value of 3. 
In contrast, 17 template AGN had L/C detections or limits well below one.
Here we have excluded Seyferts like NGC 7469, with independent evidence
for strong  circumnuclear star formation. Those systems can reach 
starburst-like PAH strength in the
integrated spectrum  (Table~2 of \cite{genzel98}). 
Figure~2 of Lutz et al. (1998) gives an impression of the 
uncertainties of the L/C ratio due to limited S/N, which may affect the 
classification in some cases.  
Several ULIRG spectra are too noisy for classification,
we include them as uncertain (`?') in Table~\ref{tab:comp}. For 
sources with available
mid-infrared fine structure line spectroscopy (\cite{genzel98}) we have 
considered the excitation of the emission line spectrum as well. Both
criteria agree well with the exception of Mrk 273 which has somewhat
low but still starburst-like PAH L/C of 1.7 and an AGN-like high excitation fine
structure line spectrum. We have chosen to classify it as AGN though it
likely represents a mixed case with both AGN and starburst contributing
strongly to the total luminosity. {\em Unique classifications were also assigned to 
other sources on the border between the two 
regimes}. While such unique classifications do not represent the fact 
that ULIRGs may host both phenomena, 
they are required for a meaningful comparison of infrared and optical
results concerning the {\em dominant} source of luminosity.

High quality optical spectra are available 
for 48 of 83 ULIRGs and IR-bright AGNs observed with ISO
(see notes to Table~\ref{tab:comp}).
Following Veilleux and Osterbrock (1987), we use
ratios of [\ion{O}{3}] $\lambda$5007/H$\beta$, 
[\ion{N}{2}] $\lambda$6583/H$\alpha$,
[\ion{S}{2}] $\lambda\lambda$6716,6731/H$\alpha$, and
[\ion{O}{1}] $\lambda$6300/H$\alpha$ to distinguish the different ionization
mechanisms of \ion{H}{2}, LINER, and Seyfert galaxies. The adopted limits of 
these three regions in the diagnostic diagrams are those shown in Figure~6 of
Veilleux et al. (1995). We used extinction corrected line ratios but 
note that, despite the typically high reddening of ULIRGs,
corrections are small due to the close proximity in wavelength of the lines 
forming the ratios. When available, we used
line ratios corrected for the presence of underlying Balmer stellar absorption
features. This consistenly applied
classification scheme caused some reclassifications with respect to the 
original papers. The nuclear optical spectral types, listed in 
Table~\ref{tab:comp}, were obtained through 
apertures of typically $\sim$2\arcsec\, corresponding to between 700pc for the
most nearby ULIRG Arp 220 and $\sim$10kpc for distant ULIRGs at redshift 
approaching 0.3. Kim et al. (1998) and Veilleux, Kim, \& Sanders (1999) used
2\arcsec\ slitwidth $\times$ 4kpc along the slit to minimize aperture size 
effects. 
These apertures include the major part or all of the active regions of ULIRGs.
{\em Again, we have adopted unique classifications}.
For double nucleus sources with optical spectroscopy for 
both nuclei, we have used average line
ratios, since both nuclei will be covered by the large
ISO apertures. These sources are labelled (D) in Table~\ref{tab:comp}.

\section{Discussion}

Our results are presented in Table~\ref{tab:comp} and
Figure~\ref{fig:isoveill}. ISO and optical classification agree well if optical 
\ion{H}{2} and LINER types are grouped together: All but one of 23 ISO 
starbursts are optical \ion{H}{2} or LINER, whereas 11 of 16 ISO AGNs are 
optical Seyferts. Reversely, all but one of the optical Seyferts of type 1 
and 2 are classified as ISO AGN. Objects with uncertain ISO classifications 
turn out to be \ion{H}{2} galaxies or LINERs, just like the ISO starbursts.
This agrees with the previous suggestion that they are starbursts, 
based on their average ISO spectrum and their infrared 
spectral energy distributions (\cite{lutz98}). 

The presence of a broad-line region (BLR) is an unambiguous AGN indicator.
Infrared spectroscopy and optical polarimetry partly circumvent the extinction
problems of optical BLR searches by observing
in a less obscured wavelength regime, or by using scattered light.
Table~\ref{tab:comp} includes a comparison to published BLR searches. 
Sources are labelled yes (no) in the respective columns if
near-infrared spectroscopy did (not) show broad components to hydrogen 
recombination lines, or if optical polarimetry has (not)
detected broad components to Balmer lines. Most entries are from
the complete far-infrared selected sample of Veilleux, Sanders, \& Kim 
(1997, 1999).
We note however that some other BLR searches 
were biased to ULIRGs already considered AGNs from optical spectroscopy.
In our sample, BLRs are found only in sources that are optically
classified as Seyfert and ISO classified as AGN, giving further
weight to the results from optical/near-infrared spectroscopy alone
(Veilleux, Sanders, \& Kim 1997, 1999).

\subsection{The nature of infrared-selected LINERs}

Galaxies optically classified as LINERs are well known to be a mixed
population. LINER type ratios
are firstly observed for LINERs in the classical sense, i.e. small low 
ionization nuclear emission regions in often fairly normal galaxies, which 
may represent the `lower end' of the AGN phenomenon
(e.g. \cite{ho97}; \cite{maoz95}; but see \cite{maoz98}), with 
photoionization likely dominating the gas excitation. 
Secondly, LINER type
ratios are observed in gas excited by ionizing shocks, as 
found in galactic winds powered by star formation in starburst galaxies like
M\,82 (\cite{chevalier85}). In such systems, LINER ratios are often found
to large distances from the nucleus.
This degeneracy has also affected optical classification of ULIRGs.
Sanders et al. (1988), in their definition of the class, assigned LINERs
to the AGN group. However, spatially resolved studies in both the optical
(Heckman, Armus, \& Miley 1987,1990; \cite{veilleux95}; \cite{kim98}) and 
soft X-ray (\cite{heckman96}) have provided evidence for large scale 
supernova-driven
galactic `superwinds' being responsible for LINER-type optical spectra
of ULIRGs and other infrared-selected galaxies. This suggested that 
selection by far-infrared flux tends
to pick a different, shock-dominated population of LINERs than optical
surveys of the nuclei of normal galaxies.   

The notion that LINER spectra of infrared-selected galaxies and ULIRGs are due 
to shocks in galactic outflows, and ultimately related to star formation, is 
strongly supported by the ISO/optical comparison: 10 of 17 optical
LINERs are put in the starburst group by the ISO spectroscopy, and only
two into the AGN group. The number of starburst LINERs rises to 15 of 17 if 
one includes the uncertain ISO classifications, which likely are starbursts.
The PAH diagnostic does not directly probe the LINER gas, but determines the
relative importance of starburst and AGN activity.
Figure~\ref{fig:isoveill} demonstrates the close similarity of
the PAH properties of optical HII and LINER types - they
exhibit equally high levels of star forming activity, and low AGN
frequency. By analogy to nearby starbursts with large-scale
LINER emission, the strong star formation activity in LINER ULIRGs will be 
able to power the LINER emission via superwinds. 
The emerging scenario for LINER ULIRGs is one of starburst activity 
extending from the
outskirts to the obscured layers probed by ISO spectroscopy, with 
optical LINER emission probing starburst-driven shocks and superwinds in 
the less obscured outer regions. It would be difficult to explain how an 
energetically dominant central AGN could
photoionize the LINER region and still remain undetected
in the mid-infrared which is much better at penetrating the
obscuring dust. 

There are also quantitative difficulties to explain LINER emission in ULIRGs
with AGN photoionization. To obtain the low ionization parameter 
$log U = log(Q_{Lyc}/4\pi R^2 n c) \sim -4$ required for a LINER 
spectrum, the emitting clouds
have to be either at large radii from a central AGN of 
ULIRG-like luminosity, or the clouds have to be very dense.
The first possibility is strongly constrained by the small sizes of the 
active regions of ULIRGs. A variety of indicators including
radio continuum (\cite{condon91}), 10$\mu$m and mm dust emission 
(\cite{soifer99}; \cite{sakamoto99}), and near-IR line emission 
(\cite{armus95}; \cite{genzel98}) suggest that most of the luminosity and 
line emission of 
a ULIRG is produced in a region of only a few hundred parsecs size. 
Scaling the properties
of the Seyfert 2 AGN of the Circinus galaxy as modelled by Moorwood et al.
(1996), Q$_{Lyc}\sim 10^{53.3}$\,photons/s and 
L$_{Bol}\sim 5\times 10^9$\,L$_{\sun}$, 
to a ULIRG luminosity results in a Lyman continuum
flux of Q$_{Lyc}\sim 10^{55\ldots56}$\,photons/s, the lower end
being a conservative estimate including uncertainties of AGN 
modelling and beaming. Adopting a density of the 
ionized gas of $\leq$1000\,cm$^{-3}$, characteristic of most of the ionized gas 
in luminous infrared galaxies, the size of a LINER photoionized region
with $log U=-4$ would have to be as large as 5\,kpc, clearly incompatible
with few hundred pc to 1\,kpc scales of the active regions of ULIRGs.

The second possibility -- high densities -- cannot fit
the low measured density-sensitive ratios of the mid-infrared [\ion{S}{3}]
18.71$\mu$m, 33.48$\mu$m lines in ULIRGs (\cite{genzel98}). These low ratios 
are interpreted as coming from gas close to the low density limit of the 
[\ion{S}{3}] ratio, i.e. at a few hundred to one thousand 
electrons cm$^{-3}$, and in addition slightly decreased by differential 
extinction. At densities in excess of 10$^5$\,cm$^{-3}$, the intrinsic ratio
is a factor of 20 higher. Matching this to the observations would imply 
extreme differential extinction between
18.71$\mu$m and 33.48$\mu$m which would require a highly improbable pure
screen extinction and make the extinction-corrected emission
incompatible with the ULIRG bolometric luminosity. Very high densities
are also incompatible with optical [\ion{S}{2}] ratios (e.g. \cite{armus89};
\cite{veilleux95}, \cite{veilleux99a}; \cite{kim98})
which, however, may preferentially probe outer regions of the ULIRGs observed.
Similar problems to match the observed [\ion{S}{3}] ratio will occur in 
LINER models photoionized by hot stars, which require high densities   
(\cite{shields92}; \cite{veilleux95}; \cite{taniguchi99}). 

A further incompatibility between photoionized LINER models and ULIRG 
observations
arises in the ratio of the [\ion{Ne}{3}]15.55$\mu$m and 
[\ion{Ne}{2}]12.81$\mu$m lines. LINER photoionization models predict this
ratio to be high, e.g. 5 in the model of Spinoglio and Malkan (1992).
In contrast, ISO-SWS observations of the two starburst-like
ULIRGs Arp~220 and NGC~6240 obtain limits or values for this ratio
substantially less than 1 (E. Sturm, priv. comm.; Egami et al., in prep.).
Only the weak fine structure line emission detected in the Seyfert~1 type
ULIRG Mrk~231 may be compatible with such a model (Rigopoulou et al., in prep.).

These quantitative considerations demonstrate the difficulty to explain
the bulk of a ULIRG's line emission with a photoionized LINER model. 
They allow, however, the simultaneous presence of a fainter AGN, especially if 
it were obscured towards our line of sight. The 
main result from ISO fine structure line spectroscopy of ULIRGs
(\cite{genzel98}), that most ULIRGs are predominantly starburst powered,
implies that any coexisting AGN would produce only a minor part to the 
bolometric luminosity.  
Such a minor AGN might still contribute to some of the extended 
emission if it were less obscured in other directions, in the spirit of
a Seyfert `ionization cone'. While this cannot
be ruled out, we note that the observed extended line 
emission in ULIRGs matches well superwind models and observations but is
quite different from the extended narrow line region of an AGN 
`ionization cone': While ionization cones tend to be high excitation
(e.g. \cite{bergeron89}; \cite{robinson94}; \cite{wilson94}), the extended 
emission in ULIRGs is of low excitation LINER type. Ionization cones
also tend to be kinematically little disturbed on larger scales, in contrast
to line splitting observed in superwinds.
 
\subsection{Can obscured AGN remain hidden?}

The presence of both starburst and AGN activity in ULIRGs naturally raises
the question of their evolutionary connection. Sanders et al. (1988) put forward
the by now classical evolutionary scenario in which interaction and merging 
of the parent galaxies trigger starburst activity that later subsides
while an AGN increasingly dominates the luminosity and expels the obscuring 
dust. The large concentrations of gas in the nuclear regions
of ULIRGs clearly play an important role in fueling both starburst and AGN
activity, but the exact relation between the two phenomena remains
difficult to model. 
Furthermore, the lack of a clear correlation between ISO spectral 
classification and the nuclear
separation (\cite{lutz98}) indicates a more complex relation than
expected for a straightforward evolution from starburst to AGN, with
local and short term conditions likely being more important.

Coexistence between a faint AGN and an energetically dominant starburst
may lead to optical AGN spectra classified as starburst by ISO. We find only
one such source. However, high spatial resolution studies indicate that 
such systems may be more common than indicated 
in our spectra which sample a large part of the active regions 
(optical) or the entire galaxies (ISO).
UGC 5101 serves as an example: Both groundbased K band imaging spectroscopy
(\cite{genzel98}) and NICMOS multicolour imaging (\cite{evans99}) indicate
a central compact AGN surrounded by a $\sim$1\,kpc radius ring
of vigorous star formation. NICMOS images suggest similar morphologies 
for some other ULIRGs (\cite{evans99}). These systems are a reminder that 
power sources in ULIRGs are not exclusive: Both AGN and 
starburst may be present in a galaxy even if one dominates the total power.

The inverse type of discrepant spectra, namely optical starbursts or LINERs
classified as AGN by ISO, are of particular interest. They correspond
to buried luminous AGN that have often been invoked as source of the luminosity
not accounted for by optical and near-infrared observations.
Such buried AGN naturally fit the 
classical evolutionary scenario, at a point where the AGN is already dominant
but has not yet expelled the obscuring dust. We do find such systems, but their
number is not large: only 5 of the 36 optical \ion{H}{2}/LINER galaxies in our 
sample. This is not 
a trivial result -- the gas and dust concentration in ULIRGs is large enough
to strongly reduce the mid-infrared and completely block the optical 
radiation if smoothly distributed as a homogeneous absorber. In such a 
scenario, galaxies would be common with just recognizable AGN in 
the mid-infrared, but optical spectra dominated by unimportant `surface' 
star formation. The small number of such systems suggests a 
different picture: Quickly after the onset of AGN activity, radiation and 
outflow from the AGN will manage to break the clumpy obscuring screen at 
least in certain directions
and become visible, though perhaps still only in attenuated or scattered light,
over a wide wavelength range. Superwinds from co-existing starbursts 
will support this process.
More deeply embedded AGN, invisible in  the mid-infrared as well as 
the optical, will not be identifiable by differing mid-infrared and optical
classifications. However, with the ULIRG starburst activity probed in the 
mid-infrared being sufficient to power the bulk of the bolometric luminosity 
(\cite{genzel98}), such deeply embedded AGN would not be energetically 
important, even if they existed.

\acknowledgements
This paper is based on studies done in 
collaboration with D.-C. Kim, A. Moorwood, D. Rigopoulou, D.B. Sanders, 
H.W.W. Spoon, E. Sturm, D. Tran, and the ZZULIRG consortium.
S. V. acknowledges the support of NASA through LTSA grant NAG~56547. 
The ISO Spectrometer Data Center at MPE is supported by DLR.

\clearpage
\renewcommand{\baselinestretch}{0.95}
\begin{table}
\caption{ISO and optical classifications}
\scriptsize
\begin{tabular}{lcccll}
\tableline
\tableline
Source                 &z    &ISO&Optical                 &BLR &BLR  \\ 
                       &     &Type&Type                   &NIR &Pol\\ 
\tableline
00183$-$7111           &0.327&AGN&LINER\tablenotemark{j,*}&    &  \\ 
00188$-$0856           &0.129&SB &LINER\tablenotemark{c}  &no\tablenotemark{n}&\\
00397$-$1312           &0.262&AGN&HII\tablenotemark{c}    &no\tablenotemark{m}&\\
00509$+$1225 IZw1      &0.061&AGN&Sy1\tablenotemark{e}    &    &  \\  
01003$-$2238           &0.118&SB &HII\tablenotemark{c}    &no\tablenotemark{m}&\\
01166$-$0844           &0.118&?  &HII\tablenotemark{c}    &    &  \\
01199$-$2307           &0.156&?  &HII\tablenotemark{c}    &    &  \\
01298$-$0744           &0.136&SB &HII\tablenotemark{c}    &    &  \\
01355$-$1814           &0.191&?  &HII\tablenotemark{c}    &    &  \\
01569$-$2939           &0.140&AGN&HII\tablenotemark{c}    &    &  \\
01572$+$0009 Mrk1014   &0.163&AGN&Sy1\tablenotemark{e}    &    &  \\
02411$+$0354 (D)       &0.144&SB &HII\tablenotemark{c}    &    &  \\
03521$+$0028           &0.152&SB &LINER\tablenotemark{c}  &no\tablenotemark{m}&\\
04103$-$2838           &0.118&?  &LINER\tablenotemark{c}  &no?\tablenotemark{n}&no\tablenotemark{r}\\
06035$-$7102 (D)       &0.079&SB &HII\tablenotemark{d}    &    &  \\
06206$-$6315           &0.092&SB &Sy2\tablenotemark{d}    &    &  \\
09104$+$4109           &0.442&AGN&Sy2\tablenotemark{f,i,*}&no\tablenotemark{p}&yes\tablenotemark{q}\\
09320$+$6134 UGC5101   &0.040&SB &LINER\tablenotemark{c}  &    &  \\
09463$+$8141           &0.155&?  &LINER\tablenotemark{c}  &    &  \\
12112$+$0305           &0.073&SB &LINER\tablenotemark{b,c}&no\tablenotemark{n}&\\
12265$+$0219 3C273     &0.158&AGN&Sy1\tablenotemark{e}    &    &  \\
12540$+$5708 Mrk231    &0.042&AGN&Sy1\tablenotemark{e}    &    &  \\
13428$+$5608 Mrk273    &0.037&AGN&Sy2\tablenotemark{b,c}  &no\tablenotemark{n}&\\
13536$+$1836 Mrk463 (D)&0.050&AGN&Sy2\tablenotemark{g}    &yes\tablenotemark{m}&yes\tablenotemark{r}\\
14348$-$1447 (D)       &0.082&SB &LINER\tablenotemark{b,c}&no\tablenotemark{m}&\\
15250$+$3609           &0.053&SB &LINER\tablenotemark{a}  &no\tablenotemark{n}&\\
15307$+$3253           &0.926&AGN&Sy2\tablenotemark{h,i}  &    &yes\tablenotemark{s}\\ 
15327$+$2340 Arp220    &0.018&SB &LINER\tablenotemark{b,c}&    &  \\ 
16333$+$4630           &0.191&?  &LINER\tablenotemark{b,c}&no\tablenotemark{m}&\\
16474$+$3430           &0.111&SB &HII\tablenotemark{b,c}  &    &  \\
16487$+$5447           &0.104&SB &LINER\tablenotemark{b,c}&    &  \\
16504$+$0228 NGC6240   &0.024&SB &LINER\tablenotemark{a}  &    &  \\
17028$+$5817 (D)       &0.106&SB &HII\tablenotemark{b,c}  &    &  \\
17068$+$4027           &0.179&AGN&HII\tablenotemark{b,c}  &no\tablenotemark{m}&\\
17179$+$5444           &0.147&AGN&Sy2\tablenotemark{c}    &yes?\tablenotemark{m}&\\
17208$-$0014           &0.043&SB &HII\tablenotemark{a}    &no\tablenotemark{n}&\\
18470$+$3233           &0.079&?  &HII\tablenotemark{b}    &    &  \\
19254$-$7245 (D)       &0.062&AGN&Sy2\tablenotemark{d}    &    &no\tablenotemark{t}\\
20100$-$4156           &0.129&SB &HII\tablenotemark{d}    &    &  \\
20551$-$4250           &0.043&SB &HII\tablenotemark{d}    &    &  \\
22491$-$1808           &0.077&SB &HII\tablenotemark{a}    &no\tablenotemark{n}&\\
23060$+$0505           &0.174&AGN&Sy2\tablenotemark{c}    &yes\tablenotemark{m,o}&\\
23129$+$2548           &0.179&AGN&LINER\tablenotemark{c}  &    &  \\
23128$-$5919 (D)       &0.045&SB &HII\tablenotemark{d}    &    &no\tablenotemark{r}\\
23230$-$6926           &0.106&?  &LINER\tablenotemark{d}  &    &  \\
23327$+$2913           &0.107&?  &LINER\tablenotemark{c}  &    &  \\
23365$+$3604           &0.064&SB &LINER\tablenotemark{a}  &no\tablenotemark{m}&\\
23389$-$6139           &0.093&SB &HII\tablenotemark{d}    &    &  \\
\tableline
\end{tabular}
\label{tab:comp}
\tablenotetext{}{Unique ISO and optical classifications were assigned even for potentially mixed systems. Optical references:
$^{a}${Veilleux et al. 1995,}
$^{b}${Kim et al. 1998,}
$^{c}${Veilleux, Kim, \& Sanders 1999,}
$^{d}${Duc et al. 1997,}
$^{e}${Sanders et al. 1988a,}
$^{f}${Kleinmann et al. 1988,}
$^{g}${Shuder \& Osterbrock 1981,}
$^{h}${Cutri et al. 1994,}
$^{i}${Evans et al. 1998,}
$^{j}${Armus et al. 1989,}
$^{*}${using observed ratios}}
\tablenotetext{}{References for broad-line region searches:
$^{m}${Veilleux, Sanders, \& Kim 1997,}
$^{n}${Veilleux, Sanders, \& Kim 1999,}
$^{o}${Hines 1991,}
$^{p}${Soifer  et al. 1996,}
$^{q}${Hines \& Wills 1993 (broad MgII),}
$^{r}${Young et al. 1996,}
$^{s}${Hines et al. 1995,}
$^{t}${Heisler et al. 1997}}
\normalsize
\end{table}

\clearpage
\renewcommand{\baselinestretch}{1.0}

\epsscale{0.53}
\plotone{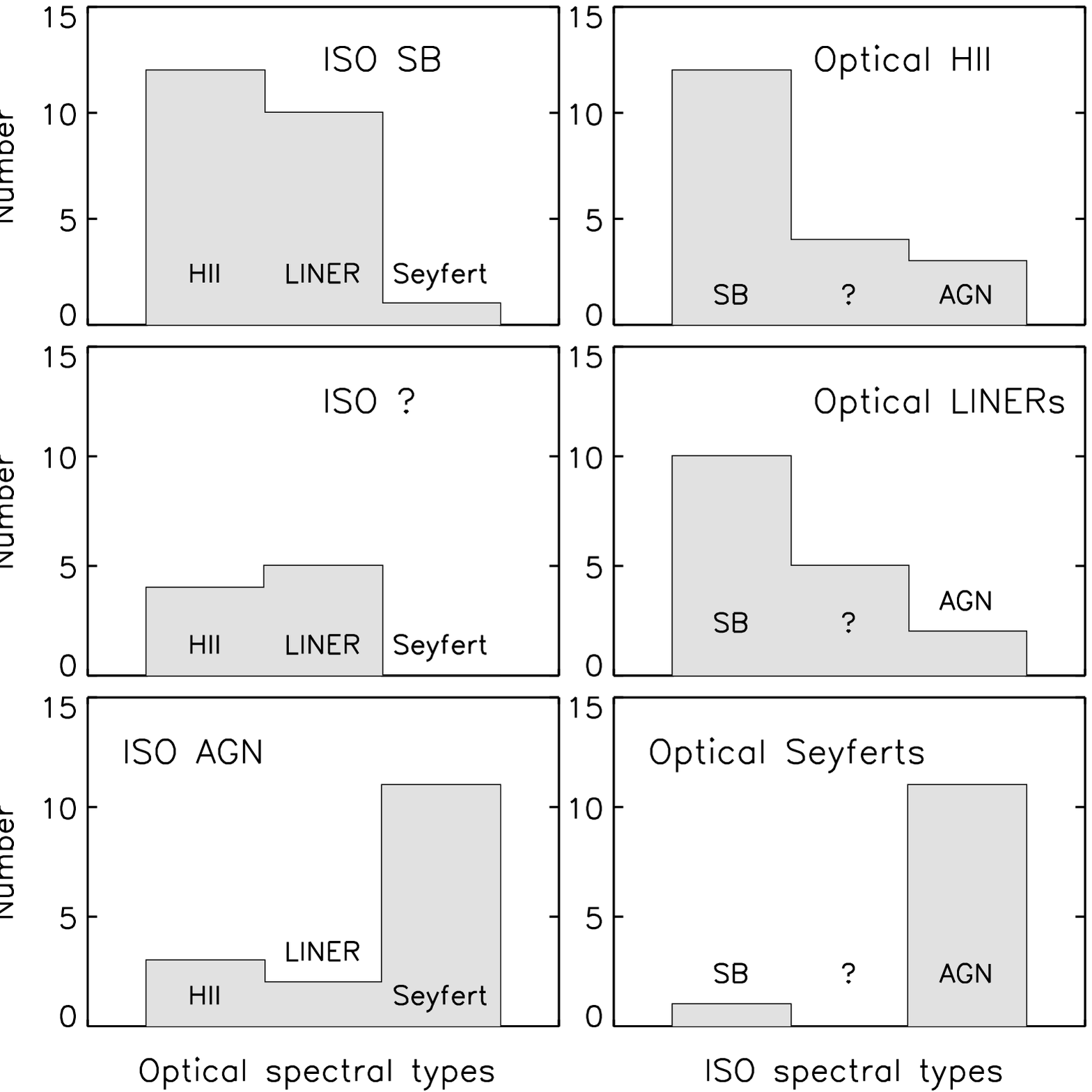}

\figcaption[pahopt1.eps]{Histograms comparing optical
and ISO spectral classifications. On the left, the distributions
of the optical spectral types are presented for ULIRGs of the different 
ISO types: 
From top to bottom starburst, uncertain due to noisy spectra, and AGN. 
On the right, distributions of the ISO spectral types are presented for 
ULIRGs of the different optical spectral types.
\label{fig:isoveill}}


\begin{thebibliography}{dum}
\bibitem[Armus, Heckman, \& Miley 1989]{armus89} Armus L., Heckman T.E., 
  \& Miley G.K. 1989, \apj, 347,727
\bibitem[Armus et al. 1995]{armus95} Armus L., Shupe D.L., Matthews K.,
  Soifer B.T., Neugebauer G. 1995a, \apj, 440, 200
\bibitem[Aussel et al. 1998]{aussel98}Aussel H., Gerin M., Boulanger F.,
  D\'esert F.X., Casoli F., Cutri R.M., \& Signore M. 1998, \aap, 334,L73
\bibitem[Bergeron et al. 1989]{bergeron89} Bergeron J., Petitjean P., Durret F.
  1989, \aap, 213, 61
\bibitem[Chevalier \& Clegg 1985]{chevalier85} Chevalier R.A, \& Clegg A.W. 
  1985, \nat, 317, 44
\bibitem[Condon et al. 1991]{condon91}Condon J.J., Huang Z.-P., Yin Q.F., 
  Thuan T.X. 1991, \apj, 378, 76
\bibitem[Cutri et al. 1994]{cutri94} Cutri R.M., Huchra J.P., Low F.J., 
  Brown R.L., \& vanden Bout P.A., 1994, \apj, 424, L65
\bibitem[Duc, Mirabel, \& Maza 1997]{duc97} Duc P.-A., Mirabel I.F., \& Maza J.
  1997, \aaps, 124, 533  
\bibitem[Evans et al. 1998]{evans98} Evans A.S., Sanders D.B., Cutri R.M.,
  Radford S.J.E., Surace J.A., Solomon P.M., Downes D., \& Kramer C. 1998,
  \apj, 506, 205 
\bibitem[Evans 1999]{evans99} Evans, A.S. 1999, \apss, submitted 
  (Proceedings of 1998 Ringberg workshop on ULIRGs)
\bibitem[Genzel et al. 1998]{genzel98} Genzel R., et al.
  1998, \apj, 498, 579
\bibitem[Goldader et al. 1995]{goldader95} Goldader J.D., Joseph R.D., 
  Doyon R., \& Sanders D.B. 1995, \apj, 444, 97
\bibitem[Heckman, Armus, \& Miley 1987]{heckman87} Heckman T.M., Armus L.,
  \& Miley G.K. 1987, \aj, 93,276 
\bibitem[Heckman, Armus, \& Miley 1990]{heckman90} Heckman T.M., Armus L.,
  \& Miley G.K. 1990, \apjs, 74,833 
\bibitem[Heckman et al. 1996]{heckman96} Heckman T.M., Dahlem M., Eales S.A.,
  Fabbiano G., \& Weaver K. 1996, \apj, 457, 616
\bibitem[Heisler, Lumsden, \& Bailey 1997]{heisler97} Heisler C.A., 
  Lumsden S.L., \& Bailey J.A. 1997, \nat, 385, 700
\bibitem[Hines 1991]{hines91} Hines D.C. 1991, \apj, 374, L9
\bibitem[Hines \& Wills 1993]{hines93} Hines D.C., \& Wills B.J. 1993, \apj, 415, 82
\bibitem[Hines et al. 1995]{hines95} Hines D.C., Schmidt G.D., Smith P.S.,
  Cutri R.M., \& Low F.J. 1995, \apj, 450, L1
\bibitem[Ho, Filippenko, \& Sargent 1997]{ho97} Ho L.C., Filippenko A.V.,
  \& Sargent W.L.W. 1997, \apj, 487, 568
\bibitem[Kim, Veilleux, \& Sanders 1998]{kim98} Kim D.-C., Veilleux S., 
  \& Sanders D.B. 1998, \apj, 508, 627
\bibitem[Kleinmann et al. 1988]{kleinmann88} Kleinmann S.G., Hamilton D., 
  Keel W.C., Wynn-Williams C.G., Eales S.A., Becklin E., \& Kuntz K.D. 1988,
  \apj, 328, 161
\bibitem[Lutz et al. 1998]{lutz98} Lutz D., Spoon H.W.W., Rigopoulou D.,
  Moorwood A.F.M., \& Genzel R. 1998, \apj, 505, L103
\bibitem[Maoz et al. 1995]{maoz95} Maoz D., Filippenko A.V., Ho L.C., Rix H.W.,
  Bahcall J.N., Schneider D.P., \& Macchetto F.D. 1995, ApJ, 440, 91
\bibitem[Maoz et al. 1998]{maoz98} Maoz D., Koratkar A., Shields J.C., 
  Ho L.C., Filippenko A.V., \& Sternberg A. 1998, AJ, 116, 55
\bibitem[Moorwood et al. 1996]{moorwood96} Moorwood A.F.M., Lutz D., Oliva E.,
  Marconi A., Netzer H., Genzel R., Sturm E., de Graauw Th. 1996, 
  \aap, 315, L109
\bibitem[Robinson et al. 1994]{robinson94} Robinson A., et al. 1994, \aap,
  291, 351
\bibitem[Sakamoto et al. 1999]{sakamoto99} Sakamoto K., Scoville N.Z., Yun M.S.,
  Crosas M., Genzel R., Tacconi L.J. 1999, ApJ, in press (astro-ph/9810325)
\bibitem[Sanders et al. 1988]{sanders88} Sanders D.B., Soifer B.T., Elias J.H.,
  Madore B.F., Matthews K., Neugebauer G., \& Scoville N.Z. 1988, \apj, 325, 74
\bibitem[Sanders et al. 1988a]{sanders88a} Sanders D.B., Soifer B.T., 
  Elias J.H., Neugebauer G., \& Matthews K. 1988a, \apj, 328, L35
\bibitem[Sanders \& Mirabel 1996]{sanders96} Sanders D.B., \& Mirabel I.F.
  1996, \araa, 34, 725
\bibitem[Shields 1992]{shields92} Shields J.C. 1992, ApJ, 399, L27
\bibitem[Shuder \& Osterbrock 1981]{shuder81} Shuder J.M., \& Osterbrock D.E.
  1981, \apj, 250, 55
\bibitem[Soifer et al. 1996]{soifer96} Soifer B.T., Neugebauer G., Armus L., 
  \& Shupe D.L. 1996, \aj, 111, 649
\bibitem[Soifer et al. 1999]{soifer99} Soifer B.T., Neugebauer G., Matthews K.,
    Becklin E.E, Ressler M., Werner M.W., Weinberger A.J., Egami E. 1999, ApJ,
    in press (astro-ph/9810120)
\bibitem[Spinoglio \& Malkan 1992]{spinoglio92} Spinoglio L., \& Malkan M.A. 
  1992, \apj, 399, 504 
\bibitem[Taniguchi et al. 1997]{taniguchi97} Taniguchi Y., Sato Y., Kawara K.,
 Murayama T., \& Mouri H. 1997, \aap, 318, L1
\bibitem[Taniguchi et al. 1999]{taniguchi99} Taniguchi Y., Yoshino A., 
  Ohyama Y., \& Nishiura S. 1999, ApJ, in press
\bibitem[Veilleux \& Osterbrock 1987]{veilleux87} Veilleux S., \& 
 Osterbrock D.E. 1987, \apjs, 63, 295
\bibitem[Veilleux et al. 1995]{veilleux95} Veilleux S., Kim D.C., Sanders D.B.,
 Mazzarella J.M., \& Soifer B.T. 1995, \apjs, 98, 171 
\bibitem[Veilleux, Sanders, \& Kim 1997]{veilleux97} Veilleux S., 
 Sanders D.B., \& Kim D.-C. 1997, \apj, 484, 92
\bibitem[Veilleux, Sanders, \& Kim 1999]{veilleux99} Veilleux S., Kim D.-C.,
 \& Sanders D.B. 1999, \apj, submitted 
\bibitem[Veilleux, Kim, \& Sanders 1999]{veilleux99a} Veilleux S., 
 Sanders D.B., \& Kim D.-C., 1999, \apj, submitted 
\bibitem[Wilson \& Tsvetanov 1994]{wilson94} Wilson A.S., Tsvetanov Z.I. 1994,
 \aj, 107, 1227
\bibitem[Young et al. 1996]{young96} Young S., Hough J.H., Efstathiou A., 
 Wills B.J., Bailey J.A., Ward M.J., \& Axon D.J. 1996, \mnras, 281, 1206
\end{thebibliography}
\end{document}